\documentstyle[12pt,epsfig,graphics,cite,amsfonts]{article}
\begin{document}\bibliographystyle{plain}\begin{titlepage}
\renewcommand{\thefootnote}{\fnsymbol{footnote}}\hfill
\begin{tabular}{l}HEPHY-PUB 855/07\\arXiv:YYMM.NNNN\\
December 2007\end{tabular}\\[1cm]\Large\begin{center}{\bf
STABILITY IN THE INSTANTANEOUS BETHE--SALPETER FORMALISM: REDUCED
EXACT-PROPAGATOR BOUND-STATE EQUATION WITH HARMONIC
INTERACTION}\\[1cm]\large{\bf Zhi-Feng LI}\\[.3cm]\normalsize
Faculty of Physics, University of Vienna,\\Boltzmanngasse 5,
A-1090 Vienna, Austria\\[1cm]\large{\bf Wolfgang
LUCHA\footnote[1]{\normalsize\ {\em E-mail address\/}:
wolfgang.lucha@oeaw.ac.at}}\\[.3cm]\normalsize Institute for High
Energy Physics,\\Austrian Academy of Sciences,\\Nikolsdorfergasse
18, A-1050 Vienna, Austria\\[1cm]\large{\bf Franz
F.~SCH\"OBERL\footnote[2]{\normalsize\ {\em E-mail address\/}:
franz.schoeberl@univie.ac.at}}\\[.3cm]\normalsize Faculty of
Physics, University of Vienna,\\Boltzmanngasse 5, A-1090 Vienna,
Austria\vfill{\normalsize\bf Abstract}\end{center}\normalsize
Several numerical investigations of the Salpeter equation with
static {\em confining\/} interactions of Lorentz-scalar type
revealed that its solutions are plagued by instabilities of
presumably Klein-paradox nature. By proving rigorously that the
energies of all predicted bound~states are part of real, entirely
discrete spectra bounded from below, these instabilities are
shown, for confining interactions of harmonic-oscillator shape, to
be absent for a ``reduced'' version of an instantaneous
Bethe--Salpeter formalism designed to generalize the Salpeter
equation towards an approximate inclusion of the {\em exact\/}
propagators of all bound-state constituents.\vspace{.5cm}

\renewcommand{\thefootnote}{\arabic{footnote}}\end{titlepage}

\section{Introduction}The presumably most well-known (and in
elementary particle physics most widely applied) three-dimensional
reduction of the (four-dimensional) Bethe--Salpeter formalism
\cite{BSE} for the description of bound states within {\em quantum
field theories\/}, such as, for example, in quantum
electrodynamics or quantum chromodynamics, is its instantaneous
approximation, derived by assuming any interaction between
bound-state constituents as static in their rest frame. The
additional assumption of free propagation of all bound-state
constituents then leads to Salpeter's equation, an {\em integral
equation\/} determining the bound-state's Salpeter amplitude
(which encodes the distribution of the bound-state constituents'
relative momenta) and its mass eigenvalue \cite{SE}. This equation
can be solved by, for instance, reduction to a set of {\em radial}
{\em relations\/} \cite{Lagae92,Olsson95,Olsson96} and subsequent
conversion to an {\em equivalent matrix eigenvalue~problem\/}
\cite{Lucha00:IBSEm=0,Lucha00:IBSE-C4,Lucha00:IBSEnzm,Lucha01:IBSEIAS}.

However, one would, of course, like to incorporate into the
formalism also effects such as dynamical breakdown of chiral
symmetry, allowing one to interpret the lowest pseudoscalar
quark--antiquark bound states as (pseudo-)Goldstone bosons. This
necessitates to take into account the exact propagators of the
bound-state constituents, a rather ambitious goal but certainly
missed by the free-propagator assumption on which the Salpeter
equation~relies.

One recent attempt in this direction has been undertaken, by two
of the present~authors (W.~L. and F.~F.~S.), in
Ref.~\cite{Lucha05:IBSEWEP}, with the implications of this
improvement for both energy levels and Salpeter amplitudes of the
bound states being (tentatively)
explored~in~Ref.~\cite{Lucha05:EQPIBSE}.

Unfortunately, numerical treatments of the Salpeter equation
\cite{SE} with (in configuration space linearly rising) {\em
confining\/} interaction observed, for a specific class of Lorentz
nature of this interaction, nasty instabilities of its solutions,
likely related to Klein's paradox
\cite{Parramore95,Parramore96,Uzzo99}.

In view of this clearly unsatisfactory state of the art, we
scrutinized, for various popular Lorentz structures, including one
suggested by B\"ohm, Joos, and Krammer (BJK hereafter)
\cite{BJK73,Gross91}, a {\em reduced\/} form
\cite{Henriques76,Jacobs87,Gara89,Gara90,Lucha92C} of Salpeter's
equation with harmonic-oscillator interactions, which allowed for
an analytical investigation of the stability problem
\cite{Lucha07:HORSE,Lucha07:StabOSS-QCD@Work07,Lucha07:SSSECI-Hadron07}.
By a rigorous stability analysis, we managed to prove all bound
states to be stable by demonstrating that their masses form {\em
real}, purely {\em discrete\/} spectra {\em bounded from below\/}
\cite{Lucha07:HORSE,Lucha07:StabOSS-QCD@Work07,Lucha07:SSSECI-Hadron07}.
Here, we extend this earlier analysis, with precisely the same
findings, to the generalized Salpeter~formalism of
Ref.~\cite{Lucha05:IBSEWEP}, where the behaviour of the full
propagators slightly complicates the discussion.

The outline of the paper is as follows. In Sec.~\ref{Sec:IBSEWEP}
we briefly recall the full {\em exact-propagator} {\em bound-state
equation\/} previously derived within the framework of the
specific instantaneous Bethe--Salpeter formalism introduced in
Ref.~\cite{Lucha05:IBSEWEP} and perform the (standard) truncation
of this bound-state equation to an exact-propagator version of the
reduced Salpeter equation. Assuming the integral kernel encoding
the interaction to be of convolution type, we reduce, in
Sec.~\ref{Sec:REEPS}, our truncated equation to a radial
eigenvalue equation for any Salpeter amplitude describing bound
states with spin-parity-charge conjugation quantum numbers
$J^{PC}=0^{-+}$ (which is the environment where all instabilities
we are concerned about should arise~first). For interactions of
harmonic-oscillator form in configuration space, any such radial
{\em integral\/} equation simplifies to an ordinary {\em
differential\/} equation, given, for various kernels, in
Sec.~\ref{Sec:CHOI}. All these differential equations are then
transformed, along the lines sketched in Sec.~\ref{Sec:TSE}, into
eigenvalue equations for Schr\"odinger operators, which can be
analyzed by standard~means. A systematic rigorous analytical
inspection of the spectral properties of all these operators,
briefly sketched in Sec.~\ref{Sec:SDSBS}, then leads us to
conclude, in Sec.~\ref{Sec:SCO}, that for reasonable behaviour of
the exact propagators of the bound-state constituents all bound
states are indeed stable.

\section{Instantaneous Bethe--Salpeter formalism for nearly exact
propagators of the bound-state constituents}\label{Sec:IBSEWEP}
Within instantaneous formulations of the Bethe--Salpeter
framework, a bound state $|{\rm B}(P)\rangle$ of momentum $P$ and
mass $M,$ composed of a fermion of mass $m_1$ and momentum $p_1$
and an antifermion of mass $m_2$ and momentum $p_2,$ represented
by Dirac field operators $\psi_1(x_1)$ and $\psi_2(x_2),$
respectively, is described in momentum space by the equal-time
Salpeter~amplitude$$\Phi(\mbox{\boldmath{$p$}})=\int{\rm
d}^3x\,\exp(-{\rm
i}\,\mbox{\boldmath{$p$}}\cdot\mbox{\boldmath{$x$}})\,\langle
0|\psi_1(0,\zeta\,\mbox{\boldmath{$x$}})\,
\bar\psi_2(0,-\eta\,\mbox{\boldmath{$x$}})|{\rm B}(P)\rangle\
,$$involving the total momentum, $P\equiv p_1+p_2,$ the relative
momentum $p\equiv\zeta\,p_1-\eta\,p_2,$ and the relative
coordinate $x\equiv x_1-x_2$ of this two-particle system, with
$\eta$ and $\zeta$ satisfying~$\eta+\zeta=1.$ (We suppress all
spinor and internal indices and all dependence on the total
momentum~$P.$)

Instantaneous approximations to the Bethe--Salpeter equation are
found by integrating the latter over the zero component, $p_0,$ of
$p.$ The four-dimensional Bethe--Salpeter equation involves two
dynamical ingredients: the {\em exact propagators\/} of both
bound-state constituents and a {\em Bethe--Salpeter kernel\/}
representing all their interactions. Let us discuss these in turn.
\begin{description}\item[Interaction kernel:]The instantaneous
approximation assumes that the Bethe--Salpeter kernel $K(p,q)$
depends, in the center-of-momentum frame of the bound state
studied, exclusively on the {\em spatial components\/},
$\mbox{\boldmath{$p$}},\mbox{\boldmath{$q$}},$ of the two relative
momenta $p,q$
involved:$$K(p,q)=K(\mbox{\boldmath{$p$}},\mbox{\boldmath{$q$}})\
.$$This regards all interactions as instantaneous and thus ignores
all retardation effects.\item[Exact propagators:]By Lorentz
covariance (if preserved by the gauge-fixing procedure), the exact
fermion propagator $S_i(p)$ is fully determined, in
parity-conserving theories, by two real $p$-dependent
Lorentz-scalar functions; the latter can be interpreted as~the
mass function $m_i(p^2)$ and wave-function renormalization factor
$Z_i(p^2)$ of the fermion:$$S_i(p)=\frac{{\rm
i}\,Z_i(p^2)}{\not\!p-m_i(p^2)+{\rm i}\,\varepsilon}\
,\quad\not\!p\equiv p^\mu\,\gamma_\mu\
,\quad\varepsilon\downarrow0\ ,\quad i=1,2\ .$$The exact
propagator $S_i(p)$ can be found as solution of the fermion
Dyson--Schwinger equation or from lattice gauge theory. The {\em
integration of the Bethe--Salpeter equation\/} over $p_0$
requires, of course, the knowledge of the explicit functional
dependence of the propagator functions $m_i(p^2)$ and $Z_i(p^2)$
on $p_0.$ In view of the lack of such information, in general
cases, these propagator functions have been assumed in
Ref.~\cite{Lucha05:IBSEWEP} to depend {\em approximately\/} just
on $\mbox{\boldmath{$p$}}$ by replacing them by $m_i(p^2)\to
m_i(\mbox{\boldmath{$p$}}^2)$ and $Z_i(p^2)\to
Z_i(\mbox{\boldmath{$p$}}^2).$ Moreover, we impose as reasonable
constraints $0<m_i(\mbox{\boldmath{$p$}}^2)<\infty$ and
$0<Z_i(\mbox{\boldmath{$p$}}^2)\le1.$
\end{description}In the free-propagator limit, $m_i(p^2)\to m_i,$
$Z_i(p^2)\to 1,$ Salpeter's equation \cite{SE} is recovered.

Defining, for particle $i=1,2,$ free-particle energy
$E_i(\mbox{\boldmath{$p$}}),$ generalized Dirac Hamiltonian
$H_i(\mbox{\boldmath{$p$}}),$ and energy projection operators
$\Lambda_i^\pm(\mbox{\boldmath{$p$}})$ for positive or negative
energies according~to\begin{eqnarray*}E_i(\mbox{\boldmath{$p$}})
&\equiv&\sqrt{\mbox{\boldmath{$p$}}^2+m_i^2(\mbox{\boldmath{$p$}}^2)}\
,\quad i=1,2\ ,\\[1ex]H_i(\mbox{\boldmath{$p$}})&\equiv&
\gamma_0\,[\mbox{\boldmath{$\gamma$}}\cdot\mbox{\boldmath{$p$}}
+m_i(\mbox{\boldmath{$p$}}^2)]\ ,\quad i=1,2\
,\\[1ex]\Lambda_i^\pm(\mbox{\boldmath{$p$}})&\equiv&
\frac{E_i(\mbox{\boldmath{$p$}})\pm H_i(\mbox{\boldmath{$p$}})}
{2\,E_i(\mbox{\boldmath{$p$}})}\ ,\quad i=1,2\ ,\end{eqnarray*}
our full-propagator instantaneous Bethe--Salpeter equation for
fermion--antifermion bound states, proposed in
Ref.~\cite{Lucha05:IBSEWEP} as generalization of Salpeter's
equation \cite{SE}, then takes
the~form\begin{eqnarray}\Phi(\mbox{\boldmath{$p$}})&=&
Z_1(\mbox{\boldmath{$p$}}_1^2)\,Z_2(\mbox{\boldmath{$p$}}_2^2)\int\frac{{\rm
d}^3q}{(2\pi)^3}\left(\frac{\Lambda_1^+(\mbox{\boldmath{$p$}}_1)\,\gamma_0\,
[K(\mbox{\boldmath{$p$}},\mbox{\boldmath{$q$}})\,
\Phi(\mbox{\boldmath{$q$}})]\,\gamma_0\,\Lambda_2^-(\mbox{\boldmath{$p$}}_2)}
{P_0-E_1(\mbox{\boldmath{$p$}}_1)-E_2(\mbox{\boldmath{$p$}}_2)}\right.\nonumber\\[1ex]
&&\hspace{22.33ex}\left.-\frac{\Lambda_1^-(\mbox{\boldmath{$p$}}_1)\,\gamma_0\,
[K(\mbox{\boldmath{$p$}},\mbox{\boldmath{$q$}})\,
\Phi(\mbox{\boldmath{$q$}})]\,\gamma_0\,\Lambda_2^+(\mbox{\boldmath{$p$}}_2)}
{P_0+E_1(\mbox{\boldmath{$p$}}_1)+E_2(\mbox{\boldmath{$p$}}_2)}\right).
\label{Eq:IBSEWEP}\end{eqnarray}Every solution satisfies the
constraint $\Lambda_1^+(\mbox{\boldmath{$p$}}_1)\,
\Phi(\mbox{\boldmath{$p$}})\,\Lambda_2^+(\mbox{\boldmath{$p$}}_2)
=\Lambda_1^-(\mbox{\boldmath{$p$}}_1)\,\Phi(\mbox{\boldmath{$p$}})\,
\Lambda_2^-(\mbox{\boldmath{$p$}}_2)=0$~\cite{Lucha05:IBSEWEP};
this entails its projector decomposition
$\Phi(\mbox{\boldmath{$p$}})=
\Lambda_1^+(\mbox{\boldmath{$p$}}_1)\,\Phi(\mbox{\boldmath{$p$}})\,\Lambda_2^-(\mbox{\boldmath{$p$}}_2)+
\Lambda_1^-(\mbox{\boldmath{$p$}}_1)\,\Phi(\mbox{\boldmath{$p$}})\,\Lambda_2^+(\mbox{\boldmath{$p$}}_2).$

Any Bethe--Salpeter interaction kernel
$K(\mbox{\boldmath{$p$}},\mbox{\boldmath{$q$}})$ can be
represented as sum of terms each of which is the product of a
Lorentz-scalar potential function with a tensor product~of some
Dirac matrices. If in each of these terms the couplings of the
bound fermions to the effective interaction involves the same
generic Dirac matrix $\Gamma,$ and if
$V_\Gamma(\mbox{\boldmath{$p$}},\mbox{\boldmath{$q$}})$ denotes
the associated potential function, the action of the kernel
$K(\mbox{\boldmath{$p$}},\mbox{\boldmath{$q$}})$ on Salpeter
amplitudes~$\Phi(\mbox{\boldmath{$p$}})$~thus reads
\begin{equation}[K(\mbox{\boldmath{$p$}},\mbox{\boldmath{$q$}})\,\Phi(\mbox{\boldmath{$q$}})]=\sum_\Gamma
V_\Gamma(\mbox{\boldmath{$p$}},\mbox{\boldmath{$q$}})\,\Gamma\,\Phi(\mbox{\boldmath{$q$}})\,\Gamma\
.\label{Eq:K}\end{equation}

First attempts to explore the consequences of introducing the
exact propagators arising in quantum chromodynamics have been
undertaken in Ref.~\cite{Lucha05:EQPIBSE}: within the
rainbow--ladder truncation scheme the Dyson--Schwinger equation
{\em suggests\/} for light-quark propagators \cite{Maris97}
$$m(\mbox{\boldmath{$p$}}^2)=\frac{a}{1+\mbox{\boldmath{$p$}}^4/b}
+m_0\ ,\quad
Z(\mbox{\boldmath{$p$}}^2)=1-\frac{c}{1+\mbox{\boldmath{$p$}}^2/d}\
,$$with $a=0.745\;{\rm GeV},$ $b=(0.744\;{\rm GeV})^4,$
$m_0=0.0055\;{\rm GeV},$ $c=0.545,$ $d=(1.85508\;{\rm GeV})^2.$

Subjecting $\Phi(\mbox{\boldmath{$p$}})$ to either of the
(equivalent) additional constraints
$\Lambda_1^-(\mbox{\boldmath{$p$}}_1)\,\Phi(\mbox{\boldmath{$p$}})=0$
or
$\Phi(\mbox{\boldmath{$p$}})\,\Lambda_2^+(\mbox{\boldmath{$p$}}_2)=0$
yields the {\em exact-propagator counterpart\/} of the reduced
Salpeter equation\begin{eqnarray}
&&\left[P_0-E_1(\mbox{\boldmath{$p$}}_1)-E_2(\mbox{\boldmath{$p$}}_2)\right]
\Phi(\mbox{\boldmath{$p$}})\nonumber\\[1ex]
&&=Z_1(\mbox{\boldmath{$p$}}_1^2)\,Z_2(\mbox{\boldmath{$p$}}_2^2)\int\frac{{\rm
d}^3q}{(2\pi)^3}\,\Lambda_1^+(\mbox{\boldmath{$p$}}_1)\,\gamma_0\,
[K(\mbox{\boldmath{$p$}},\mbox{\boldmath{$q$}})\,\Phi(\mbox{\boldmath{$q$}})]
\,\gamma_0\,\Lambda_2^-(\mbox{\boldmath{$p$}}_2)\
.\label{Eq:RIBSEWEP}\end{eqnarray}For the study of the spectrum of
bound-state mass eigenvalues $M$ it is sufficient to consider the
center-of-momentum frame of the two-particle system defined by
$\mbox{\boldmath{$P$}}={\bf 0},$ which implies
$\mbox{\boldmath{$p$}}=
\mbox{\boldmath{$p$}}_1=-\mbox{\boldmath{$p$}}_2.$ There the time
component, $P_0,$ of the total momentum $P$ reduces to $M,$ i.e.,
$P_0=M.$ Accordingly, we will perform our spectral analysis in the
bound state's~rest~frame.

For kernels of the form (\ref{Eq:K}), by suitable generalization
of Eq.~(18) of Ref.~\cite{Olsson96}, all solutions of our
exact-propagator reduced instantaneous Bethe--Salpeter equation
(\ref{Eq:RIBSEWEP}) have to satisfy
\begin{eqnarray*}&&M\int\frac{{\rm d}^3p}{(2\pi)^3}\,{\rm
Tr}\left[\Phi^\dag(\mbox{\boldmath{$p$}})\,\Phi(\mbox{\boldmath{$p$}})\right]
=\int\frac{{\rm
d}^3p}{(2\pi)^3}\left[E_1(\mbox{\boldmath{$p$}})+E_2(\mbox{\boldmath{$p$}})\right]{\rm
Tr}\left[\Phi^\dag(\mbox{\boldmath{$p$}})\,\Phi(\mbox{\boldmath{$p$}})\right]\\[1ex]
&&+\int\frac{{\rm d}^3p}{(2\pi)^3}\int\frac{{\rm
d}^3q}{(2\pi)^3}\,Z_1(\mbox{\boldmath{$p$}}^2)\,Z_2(\mbox{\boldmath{$p$}}^2)\sum_\Gamma
V_\Gamma(\mbox{\boldmath{$p$}},\mbox{\boldmath{$q$}})\,{\rm
Tr}\left[\Phi^\dag(\mbox{\boldmath{$p$}})\,\gamma_0\,\Gamma\,
\Phi(\mbox{\boldmath{$q$}})\,\Gamma\,\gamma_0\right].\end{eqnarray*}
Recalling our line of argument given in Sec.~7 of
Ref.~\cite{Lucha07:HORSE} (see also
Ref.~\cite{Lucha07:StabOSS-QCD@Work07}), in this~relation both the
integral on its left-hand side and the first term on its
right-hand side are obviously nonvanishing and real while the
second term on its right-hand side is real if
$Z_1(\mbox{\boldmath{$p$}}^2)\,Z_2(\mbox{\boldmath{$p$}}^2)$ is
real, the potential functions
$V_\Gamma(\mbox{\boldmath{$p$}},\mbox{\boldmath{$q$}})$ satisfy
$V^\ast_\Gamma(\mbox{\boldmath{$q$}},\mbox{\boldmath{$p$}})=
V_\Gamma(\mbox{\boldmath{$p$}},\mbox{\boldmath{$q$}}),$ and the
Dirac couplings $\Gamma$ satisfy
$\gamma_0\,\Gamma^\dag\,\gamma_0=\pm\Gamma.$ If this holds, all
bound-state mass eigenvalues $M$ are necessarily~real.

\section{Radial eigenvalue equations for pseudoscalar
states}\label{Sec:REEPS}Following, or mimicking, the path paved in
Refs.~\cite{Lagae92,Olsson95,Olsson96,Lucha07:HORSE,
Lucha07:StabOSS-QCD@Work07,Lucha07:SSSECI-Hadron07}, as first step
of our analysis we simplify the bound-state equation
(\ref{Eq:RIBSEWEP}), for given Dirac structures
$\Gamma\otimes\Gamma$ of the interaction, to radial eigenvalue
equations by factorizing off all dependence on angular variables,
which~for Bethe--Salpeter interaction kernels of {\em convolution
type\/}, $K(\mbox{\boldmath{$p$}},\mbox{\boldmath{$q$}})
=K(\mbox{\boldmath{$p$}}-\mbox{\boldmath{$q$}}),$ is a
trivial~one.

For notational brevity we restrict the presentation of our
considerations to bound states built up by fermion and associated
antifermion. This entails, with $p\equiv|\mbox{\boldmath{$p$}}|,$
for the masses of both bound-state constituents
$m_1(\mbox{\boldmath{$p$}}_1^2)=m_2(\mbox{\boldmath{$p$}}_2^2)=:m(p),$
for their renormalization factors
$Z_1(\mbox{\boldmath{$p$}}_1^2)=Z_2(\mbox{\boldmath{$p$}}_2^2)=:Z(p)$
and for their energies
$E_1(\mbox{\boldmath{$p$}}_1)=E_2(\mbox{\boldmath{$p$}}_2)=:E(p)
\equiv\sqrt{p^2+m^2(p)}.$

On simple and purely energetic grounds, instabilities of the kind
we worry about~should manifest themselves first for pseudoscalar
bound states \cite{Parramore96}. Consequently, we will consider
fermion--antifermion bound states with total spin $J,$ parity
$P=(-1)^{J+1}$ and (well-defined) charge-conjugation quantum
number $C=(-1)^J.$ The particular projector structure of~the
bound-state equation (\ref{Eq:RIBSEWEP}) entails, for all its
solutions $\Phi(\mbox{\boldmath{$p$}}),$ the unique component
structure $\Phi(\mbox{\boldmath{$p$}})=
\Lambda_1^+(\mbox{\boldmath{$p$}}_1)\,\Phi(\mbox{\boldmath{$p$}})\,
\Lambda_2^-(\mbox{\boldmath{$p$}}_2)$ \cite{Lucha07:HORSE}. As
consequence of this, for the states under consideration any
solution $\Phi(\mbox{\boldmath{$p$}})$ of Eq.~(\ref{Eq:RIBSEWEP})
involves only one independent component,
$\phi(\mbox{\boldmath{$p$}}).$ Dropping~the indices $i=1,2$ in the
definitions of Sec.~\ref{Sec:IBSEWEP}, any generic solution of
Eq.~(\ref{Eq:RIBSEWEP}) is thus of the~form
$$\Phi(\mbox{\boldmath{$p$}})=\phi(\mbox{\boldmath{$p$}})\,
\frac{H(\mbox{\boldmath{$p$}})+E(\mbox{\boldmath{$p$}})}
{E(\mbox{\boldmath{$p$}})}\,\gamma_5\equiv2\,
\phi(\mbox{\boldmath{$p$}})\,\Lambda^+(\mbox{\boldmath{$p$}})\,
\gamma_5\ .$$The bound states in the focus of our interest, i.e.,
the pseudoscalar states, are characterized by total spin $J=0$ and
thus by the spin-parity-charge conjugation assignment
$J^{PC}=0^{-+}.$

Stripping off all spherical harmonics reduces
Eq.~(\ref{Eq:RIBSEWEP}) to an equation for the radial~factor,
$\phi(p),$ in the independent amplitude
$\phi(\mbox{\boldmath{$p$}}).$ Therein the interaction between the
bound-state constituents defined, in configuration space, by some
spherically symmetric static potential $V(r),$
$r\equiv|\mbox{\boldmath{$x$}}|,$ enters in form of a set of
Fourier--Bessel transforms $V_L(p,q)$
($L=0,1,2,\dots$):$$V_L(p,q)\equiv 8\pi\int\limits_0^\infty{\rm
d}r\,r^2\,j_L(p\,r)\,j_L(q\,r)\,V(r)\ ,\quad L=0,1,2,\dots\ ,$$
where $j_n(z),$ for $n=0,\pm1,\pm2,\dots,$ label the spherical
Bessel functions of the first kind \cite{Abramowitz}. Specifying
the Lorentz behaviour of the Bethe--Salpeter kernel
$K(\mbox{\boldmath{$p$}}-\mbox{\boldmath{$q$}}),$ we thus obtain
the radial eigenvalue equations, for interactions of
Lorentz-scalar Dirac structure, $\Gamma\otimes\Gamma=1\otimes1,$
$$2\,E(p)\,\phi(p)-\frac{1}{2}\,Z^2(p)\int\limits_0^\infty\frac{{\rm
d}q\,q^2}{(2\pi)^2}\left[\left(1+\frac{m(p)\,m(q)}{E(p)\,E(q)}\right)V_0(p,q)
-\frac{p\,q\,V_1(p,q)}{E(p)\,E(q)}\right]\phi(q)=M\,\phi(p)\ ,$$
for interactions of time-component Lorentz-vector Dirac structure,
$\Gamma\otimes\Gamma=\gamma^0\otimes\gamma^0,$
$$2\,E(p)\,\phi(p)+\frac{1}{2}\,Z^2(p)\int\limits_0^\infty\frac{{\rm
d}q\,q^2}{(2\pi)^2}\left[\left(1+\frac{m(p)\,m(q)}{E(p)\,E(q)}\right)V_0(p,q)
+\frac{p\,q\,V_1(p,q)}{E(p)\,E(q)}\right]\phi(q)=M\,\phi(p)\ ,$$
for interactions of Lorentz-vector Dirac structure,
$\Gamma\otimes\Gamma=\gamma_\mu\otimes\gamma^\mu,$
$$2\,E(p)\,\phi(p)+Z^2(p)\int\limits_0^\infty\frac{{\rm
d}q\,q^2}{(2\pi)^2}\left(2-\frac{m(p)\,m(q)}{E(p)\,E(q)}\right)V_0(p,q)\,
\phi(q)=M\,\phi(p)\ ,$$for interactions of Lorentz-pseudoscalar
Dirac structure, $\Gamma\otimes\Gamma=\gamma_5\otimes\gamma_5,$
$$2\,E(p)\,\phi(p)-\frac{1}{2}\,Z^2(p)\int\limits_0^\infty\frac{{\rm
d}q\,q^2}{(2\pi)^2}\left[\left(1-\frac{m(p)\,m(q)}{E(p)\,E(q)}\right)V_0(p,q)
-\frac{p\,q\,V_1(p,q)}{E(p)\,E(q)}\right]\phi(q)=M\,\phi(p)\ ,$$
and, for interactions of BJK \cite{BJK73,Gross91} Dirac structure,
$\Gamma\otimes\Gamma=\frac{1}{2}\,
(\gamma_\mu\otimes\gamma^\mu+\gamma_5\otimes\gamma_5-1\otimes1),$
$$2\,E(p)\,\phi(p)+Z^2(p)\int\limits_0^\infty\frac{{\rm
d}q\,q^2}{(2\pi)^2}\,V_0(p,q)\,\phi(q)=M\,\phi(p)\ .$$

\section{Confining interactions of harmonic-oscillator
type}\label{Sec:CHOI}For pure harmonic-oscillator interactions,
represented by the configuration-space potential$$V(r)=a\,r^2\
,\quad a=a^\ast\ne0\ ,\quad r\equiv |\mbox{\boldmath{$x$}}|\ ,$$
upon trading the harmonic-oscillator interaction for the
second-order differential operators$$D_p^{(L)}\equiv\frac{{\rm
d}^2}{{\rm d}p^2}+\frac{2}{p}\,\frac{{\rm d}}{{\rm
d}p}-\frac{L\,(L+1)}{p^2}\ ,\quad L=0,1,2,\dots\ ,$$(which are
nothing but the Laplacian
$\Delta\equiv\mbox{\boldmath{$\nabla$}}\cdot\mbox{\boldmath{$\nabla$}}$
acting on states of angular momentum~$L$), the Fourier--Bessel
{\em integral\/} transforms $V_L(p,q)$ encoding all interactions
explicitly read~\cite{Lucha07:HORSE}
\begin{equation}V_L(p,q)=-\frac{(2\pi)^2\,a}{q^2}\,D_p^{(L)}\,\delta(p-q)\
,\quad L=0,1,2,\dots\ .\label{Eq:HOI}\end{equation}In this case
all integral equations representing our reduced exact-propagator
instantaneous bound-state equation (\ref{Eq:RIBSEWEP}) simplify to
{\em second-order homogeneous linear ordinary differential
equations}; the latter read, for interactions of Lorentz-scalar
Dirac structure,~\mbox{$\Gamma\otimes\Gamma=1\otimes1,$}
\begin{equation}\left[2\,E(p)+\frac{Z^2(p)\,a}{2}
\left(D_p^{(0)}+\frac{m(p)}{E(p)}\,D_p^{(0)}\,\frac{m(p)}{E(p)}
-\frac{p}{E(p)}\,D_p^{(1)}\,\frac{p}{E(p)}\right)\right]\phi(p)=M\,\phi(p)\
,\label{Eq:ODE-LS}\end{equation}for interactions of time-component
Lorentz-vector Dirac structure,
$\Gamma\otimes\Gamma=\gamma^0\otimes\gamma^0,$
\begin{equation}\left[2\,E(p)-\frac{Z^2(p)\,a}{2}
\left(D_p^{(0)}+\frac{m(p)}{E(p)}\,D_p^{(0)}\,\frac{m(p)}{E(p)}
+\frac{p}{E(p)}\,D_p^{(1)}\,\frac{p}{E(p)}\right)\right]\phi(p)=M\,\phi(p)\
,\label{Eq:ODE-TCLV}\end{equation}for interactions of
Lorentz-vector Dirac structure,
$\Gamma\otimes\Gamma=\gamma_\mu\otimes\gamma^\mu,$
\begin{equation}\left[2\,E(p)-Z^2(p)\,a
\left(2\,D_p^{(0)}-\frac{m(p)}{E(p)}\,D_p^{(0)}\,\frac{m(p)}{E(p)}
\right)\right]\phi(p)=M\,\phi(p)\
,\label{Eq:ODE-LV}\end{equation}for interactions of
Lorentz-pseudoscalar Dirac structure,
$\Gamma\otimes\Gamma=\gamma_5\otimes\gamma_5,$
\begin{equation}\left[2\,E(p)+\frac{Z^2(p)\,a}{2}
\left(D_p^{(0)}-\frac{m(p)}{E(p)}\,D_p^{(0)}\,\frac{m(p)}{E(p)}
-\frac{p}{E(p)}\,D_p^{(1)}\,\frac{p}{E(p)}\right)\right]\phi(p)=M\,\phi(p)\
,\label{Eq:ODE-LPS}\end{equation}and, for interactions of BJK
\cite{BJK73,Gross91} Dirac structure,
$\Gamma\otimes\Gamma=\frac{1}{2}\,
(\gamma_\mu\otimes\gamma^\mu+\gamma_5\otimes\gamma_5-1\otimes1),$
\begin{equation}\left[2\,E(p)-Z^2(p)\,a\,D_p^{(0)}\right]\phi(p)=M\,\phi(p)\
.\label{Eq:ODE-BJK}\end{equation}Apart from the
Lorentz-pseudoscalar case all differential operators on the
left-hand sides of these eigenvalue equations are {\em not\/}
self-adjoint, their {\em spectra\/}, therefore, not necessarily
real. Nevertheless, by our arguments of Sec.~\ref{Sec:IBSEWEP} and
the fact that our potential
function~$V_\Gamma(\mbox{\boldmath{$p$}},\mbox{\boldmath{$q$}})$~is
the Fourier transform
$V_\Gamma(\mbox{\boldmath{$p$}},\mbox{\boldmath{$q$}})
=V_\Gamma(\mbox{\boldmath{$p$}}-\mbox{\boldmath{$q$}})\equiv\int{\rm
d}^3p\exp[-{\rm
i}\,(\mbox{\boldmath{$p$}}-\mbox{\boldmath{$q$}})\cdot\mbox{\boldmath{$x$}}]
\,V_\Gamma(r)$ of a real~central configuration-space potential
$V_\Gamma(r)=V^\ast_\Gamma(r)$ we can be sure that all {\em
eigenvalues\/} $M$ are~real. The above (differential) equations
may be further simplified by application of the
identities\begin{eqnarray}
D_p^{(0)}\,\frac{m(p)}{E(p)}&=&\frac{m(p)}{E(p)}\,D_p^{(0)}
+\frac{2}{E(p)}\left[\frac{{\rm d}m}{{\rm d}p}(p)
-\frac{m(p)}{E(p)}\,\frac{{\rm d}E}{{\rm d}p}(p)\right]\frac{{\rm
d}}{{\rm d}p}+\left[D_p^{(0)}\,\frac{m(p)}{E(p)}\right],
\nonumber\\[1ex]
D_p^{(0)}\,\frac{p}{E(p)}&=&\frac{p}{E(p)}\,D_p^{(0)}
+\frac{2}{E(p)}\left[1-\frac{p}{E(p)}\,\frac{{\rm d}E}{{\rm
d}p}(p)\right]\frac{{\rm d}}{{\rm
d}p}+\left[D_p^{(0)}\,\frac{p}{E(p)}\right].\label{Eq:RDR-ids}
\end{eqnarray}First of all, by adopting these identities, the
definition $E^2(p)\equiv p^2+m^2(p),$ and the~relation
$$p+m(p)\,\frac{{\rm d}m}{{\rm d}p}(p)= E(p)\,\frac{{\rm d}E}{{\rm
d}p}(p)\ ,$$it is very straightforward to convince oneself that,
in spite of its appearance, the differential operators cancel in
Eq.~(\ref{Eq:ODE-LPS}). Thus, as was the case already for the
reduced Salpeter equation \cite[Sec.~6]{Lucha07:HORSE}, for all
harmonic-oscillator interactions of Lorentz-pseudoscalar Dirac
structure $\Gamma\otimes\Gamma=\gamma_5\otimes\gamma_5$ the
eigenvalue problem of Eq.~(\ref{Eq:ODE-LPS}) is posed by a pure
multiplication~operator,$$\left\{2\,E(p)+\frac{Z^2(p)\,a}{2\,E^4(p)}
\left[2\,E^2(p)+\left(m(p)-p\,\frac{{\rm d}m}{{\rm
d}p}(p)\right)^2\right]\right\}\phi(p)=M\,\phi(p)\ ,$$which has a
continuous spectrum but no eigenvalue. Accordingly, the
Lorentz-pseudoscalar interaction kernel cannot describe bound
states and does not need to be considered further.

\section{Transformation to Schr\"odinger eigenvalue
equation}\label{Sec:TSE}Because of the momentum dependence of
$Z(p),$ $m(p),$ and $E(p),$ all our genuine differential equations
(\ref{Eq:ODE-LS}), (\ref{Eq:ODE-TCLV}), (\ref{Eq:ODE-LV}), and
(\ref{Eq:ODE-BJK}) are not standard Schr\"odinger equations.
However, they~may be easily reformulated
\cite{Lucha07:HORSE,Lucha07:StabOSS-QCD@Work07,Lucha07:SSSECI-Hadron07}
as usual Schr\"odinger eigenvalue equation for zero eigenvalue,
\begin{equation}[-D_p^{(0)}+U(p)]\,\psi(p) \equiv\left[-\frac{{\rm
d}^2}{{\rm d}p^2}-\frac{2}{p}\,\frac{{\rm d}}{{\rm
d}p}+U(p)\right]\psi(p)=0\
,\label{Eq:EV0SchrEq}\end{equation}where $U(p)$ is an auxiliary
potential to be found case by case and the Laplacian expected in a
Schr\"odinger equation is assumed to act on states of vanishing
orbital angular momentum: First, dividing by $[Z^2(p)\,a]\ne0$
(which is nonvanishing by assumption) and working~out the
derivatives with the aid of Eqs.~(\ref{Eq:RDR-ids}) simplifies all
differential equations to the common~form
\begin{equation}\left[-\frac{{\rm d}^2}{{\rm d}p^2}-2\,g(p)\,\frac{{\rm
d}}{{\rm d}p}+h(p)\right]\phi(p)=0\ ,\label{Eq:2ODE}\end{equation}
where in each case the two functions $g(p)$ and $h(p)$ may be
easily read off from Eqs.~(\ref{Eq:ODE-LS}), (\ref{Eq:ODE-TCLV}),
(\ref{Eq:ODE-LV}), or (\ref{Eq:ODE-BJK}). Here, merely $h(p)$
involves $M$ as a parameter, whereas $g(p)$ is independent of $M;$
this observation will considerably facilitate our analysis. Then,
performing the substitution $\phi(p)=f(p)\,\psi(p)$ of the
bound-state amplitude leads to the desired Schr\"odinger shape
(\ref{Eq:EV0SchrEq}), provided the transforming function $f(p)$ is
found as the solution of the differential~equation
\begin{equation}\left[\frac{{\rm d}}{{\rm d}p}+g(p)\right]f(p)
=\frac{f(p)}{p}\ ,\label{Eq:DEg}\end{equation}which may be easily
integrated, yielding the (formal) solution, up to an irrelevant
constant,\begin{equation}f(p)=p\exp\left[-\int{\rm
d}p\,g(p)\right].\label{Eq:TFf}\end{equation}Our auxiliary
potential $U(p)$ is, of course, fully determined by the quantities
$g(p)$ and~$h(p)$:\begin{equation}U(p)\equiv h(p)+\frac{{\rm
d}g}{{\rm d}p}(p)+g^2(p)\ .\label{Eq:EP}\end{equation}Clearly, if
$g(p)=1/p,$ as is the case for the time-component Lorentz-vector
Dirac structure $\Gamma\otimes\Gamma=\gamma^0\otimes\gamma^0$ and
the BJK \cite{BJK73,Gross91} Lorentz structure
$\Gamma\otimes\Gamma=\frac{1}{2}\,
(\gamma_\mu\otimes\gamma^\mu+\gamma_5\otimes\gamma_5-1\otimes1),$
the differential equation (\ref{Eq:2ODE}) is already of the
desired Schr\"odinger form (\ref{Eq:EV0SchrEq}). In these~cases,
the integration (\ref{Eq:TFf}) of our definition (\ref{Eq:DEg}) of
$f(p)$ trivially yields $f(p)=1,$ that is, $\psi(p)=\phi(p),$ and
the effective potential (\ref{Eq:EP}) becomes just the
$M$-dependent function $h(p)$: $U(p)=h(p).$ In both our nontrivial
cases, $g(p)$ reads, for the Lorentz-scalar Dirac structure
$\Gamma\otimes\Gamma=1\otimes1,$$$g(p)=\frac{1}{p}-\frac{p}
{m(p)\,E^2(p)}\left(m(p)-p\,\frac{{\rm d}m}{{\rm d}p}(p)\right)
=\frac{1}{p\,m(p)\,E^2(p)}\left(m^3(p)+p^3\,\frac{{\rm d}m}{{\rm
d}p}(p)\right)$$and, for the Lorentz-vector Dirac structure,
$\Gamma\otimes\Gamma=\gamma_\mu\otimes\gamma^\mu,$$$g(p)=\frac{1}{p}
+\frac{p\,m(p)}{E^2(p)\,[E^2(p)+p^2]}\left(m(p)-p\,\frac{{\rm
d}m}{{\rm d}p}(p)\right).$$The effective potentials $U(p)$ are,
for kernels of Lorentz-scalar Dirac structure
$\Gamma\otimes\Gamma=1\otimes1,$
$$U(p)=-\frac{2\,E(p)-M}{Z^2(p)\,a}\,\frac{E^2(p)}{m^2(p)}-
\frac{1}{2\,m^2(p)\,E^2(p)}\left[2\,E^2(p)+\left(m(p)-p\,
\frac{{\rm d}m}{{\rm d}p}(p)\right)^2\right],$$for kernels of
time-component Lorentz-vector Dirac structure
$\Gamma\otimes\Gamma=\gamma^0\otimes\gamma^0,$
$$U(p)=\frac{2\,E(p)-M}{Z^2(p)\,a}+\frac{1}{2\,E^4(p)}
\left[2\,E^2(p)+\left(m(p)-p\,\frac{{\rm d}m}{{\rm
d}p}(p)\right)^2\right],$$for kernels of Lorentz-vector Dirac
structure $\Gamma\otimes\Gamma=\gamma_\mu\otimes\gamma^\mu,$
$$U(p)=\frac{2\,E(p)-M}{Z^2(p)\,a}\,\frac{E^2(p)}{E^2(p)+p^2}-
\frac{2\,p^2}{E^2(p)\,[E^2(p)+p^2]^2}\left(m(p)-p\,\frac{{\rm
d}m}{{\rm d}p}(p)\right)^2\ ,$$and, for kernels of the BJK
\cite{BJK73,Gross91} Dirac structure
$\Gamma\otimes\Gamma=\frac{1}{2}\,
(\gamma_\mu\otimes\gamma^\mu+\gamma_5\otimes\gamma_5-1\otimes1),$
$$U(p)=\frac{2\,E(p)-M}{Z^2(p)\,a}\ .$$Interestingly, precisely a
linear dependence $m(p)\propto p$ of the mass function $m(p)$ on
$p$~entails$$m(p)-p\,\frac{{\rm d}m}{{\rm d}p}(p)=0\ .$$In this
latter case, our transformation becomes trivial also for
Lorentz-scalar ($\Gamma\otimes\Gamma=1\otimes1$) and
Lorentz-vector ($\Gamma\otimes\Gamma=\gamma_\mu\otimes\gamma^\mu$)
kernels: $g(p)=1/p$ implies $f(p)=1$ and $\psi(p)=\phi(p).$ In the
``Salpeter limit'' of {\em free\/} propagators involving {\em
constant\/} constituent masses, that~is,~if$$Z(p)\equiv1\ ,\quad
m(p)=\mbox{const}\quad\Leftrightarrow\quad\frac{{\rm d}m}{{\rm
d}p}(p)=0\ ,$$all these potentials $U(p)$ must reduce to the
corresponding expressions in Sec.~6 of Ref.~\cite{Lucha07:HORSE}.

\section{Spectra: discreteness, semiboundedness, stability}
\label{Sec:SDSBS}In the preceding section, we succeeded to rewrite
the differential equations (\ref{Eq:ODE-LS}), (\ref{Eq:ODE-TCLV}),
(\ref{Eq:ODE-LV}), and (\ref{Eq:ODE-BJK}) as eigenvalue equations
for eigenvalue zero of Schr\"odinger Hamiltonian operators of the
form ${\cal H}\equiv-\Delta+U$ (acting only on states of vanishing
orbital angular momentum).~In~order to proceed with our spectral
analysis of the bound-state masses $M,$ we recall a fundamental
theorem about the spectra of Hamiltonians with potentials
increasing beyond bound \cite{RS4}: a Schr\"odinger operator
$H\equiv-\Delta+V,$ defined as sum of quadratic forms, with
positive, locally bounded, infinitely rising potential
$V(x)\to\infty$ for $|x|\to\infty$ has a purely discrete spectrum.
This theorem may be trivially generalized to all potentials $V$
that are bounded~from~below. Thus, if the effective potential
$U(p)$ satisfies all requirements of this theorem, the spectrum of
the corresponding $M$-dependent Hamiltonian ${\cal H}$ will be,
for any value of the bound-state mass $M,$ entirely discrete: it
will consist exclusively of isolated eigenvalues ${\cal
E}_i(M)$~($i\in{\mathbb{Z}}$)~of finite multiplicity, depending,
of course, on one parameter, $M.$ By construction, the zeros of
these eigenvalue {\em functions\/} ${\cal E}_i(M)$ define the
wanted set of bound-state mass eigenvalues~$M.$

A closer inspection reveals that for sufficiently well-behaved
propagator functions $m(p)$ and $Z(p)$ [in particular, if $m(p)$
is an element of the space of differentiable functions on
${\mathbb{R}}^+$] and for an ``appropriate'' choice of the sign of
the harmonic-oscillator interaction strength $a$ all auxiliary
potentials $U(p)$ resulting, by means of Eq.~(\ref{Eq:EP}), from
the differential equations (\ref{Eq:ODE-LS}), (\ref{Eq:ODE-TCLV}),
(\ref{Eq:ODE-LV}), and (\ref{Eq:ODE-BJK}) satisfy all the
assumptions of the ``infinitely-rising-potential theorem:''
\begin{enumerate}\item The behaviour of all potentials $U(p)$ for
large relative momenta $p$ is dominated by the {\em contribution
of the kinetic part\/} $2\,E(p)$ of our exact-propagator reduced
instantaneous Bethe--Salpeter equation (\ref{Eq:RIBSEWEP}). This
contribution is necessarily proportional to $1/a$: for the choice
$a<0$ in the case of a kernel of Lorentz-scalar Dirac structure,
$\Gamma\otimes\Gamma=1\otimes1,$ and for the choice $a>0$ in the
case of interactions of time-component Lorentz-vector Dirac
structure, $\Gamma\otimes\Gamma=\gamma^0\otimes\gamma^0,$ or
Lorentz-vector Dirac structure,
$\Gamma\otimes\Gamma=\gamma_\mu\otimes\gamma^\mu,$~or BJK
\cite{BJK73,Gross91} Dirac structure,
$\Gamma\otimes\Gamma=\frac{1}{2}\,
(\gamma_\mu\otimes\gamma^\mu+\gamma_5\otimes\gamma_5-1\otimes1),$
all our potentials~$U(p)$ exhibit in the large-$p$ limit the rise
to positive infinity required by the above theorem.\item In order
to be on the safe side, we avoid the vanishing of denominators by
relying on a strict positivity $m(p)>0$ of the mass functions
$m(p)$ of the bound-state constituents: $m(p)\ne0$ for all
$p\in{\mathbb{R}}^+\equiv[0,\infty)$ should, together with
$Z(p)\ne0,$ suffice to guarantee the absence of singularities in
all our potentials $U(p)$ and, as immediate consequence, for all
potentials $U(p)$ both their {\em local boundedness\/} and their
{\em boundedness from below}. In particular instances, this
requirement of strict positivity of the mass function~$m(p)$ can
be loosened to an extent which depends on the Dirac structure
$\Gamma\otimes\Gamma$ of the~kernel: for the BJK
\cite{BJK73,Gross91} structure $\Gamma\otimes\Gamma=\frac{1}{2}\,
(\gamma_\mu\otimes\gamma^\mu+\gamma_5\otimes\gamma_5-1\otimes1),$
all prerequisites of the above theorem are satisfied
automatically, without imposing any constraint on $m(p);$ in the
case of the time-component Lorentz-vector structure
$\Gamma\otimes\Gamma=\gamma^0\otimes\gamma^0,$ we find as
sufficient to require that $m(p)$ is nonvanishing at the origin,
i.e., to demand~$m(0)\ne0;$ in the case of the Lorentz-vector
structure $\Gamma\otimes\Gamma=\gamma_\mu\otimes\gamma^\mu$ the
mass $m(p)$ even~may~be allowed to approach zero in the limit
$p\to0,$ without doing any harm, if it behaves for small $p$ like
$m(p)\propto p^d$ with an exponent
$d\in(0,\frac{1}{2}]\cup\{1\}\cup [2,\infty),$ i.e.,
$d\notin(\frac{1}{2},1)\cup(1,2).$\end{enumerate}Hence, we know
that all eigenvalues ${\cal E}_i(M)$ of all our auxiliary
Hamiltonians ${\cal H}$ are discrete.

The discreteness of all auxiliary eigenvalues ${\cal E}_i(M)$ for
all $M$ guarantees the discreteness of the spectrum of bound-state
masses $M$ \cite{Lucha07:HORSE,Lucha07:StabOSS-QCD@Work07} if for
each eigenvalue ${\cal E}_i(M)$ the derivative of ${\cal E}_i(M)$
with respect to $M$ can be shown to be strictly definite, that is,
if, for every $i,$~either$$\frac{{\rm d}{\cal E}_i}{{\rm
d}M}(M)>0\quad\forall\ M$$or$$\frac{{\rm d}{\cal E}_i}{{\rm
d}M}(M)<0\quad\forall\ M$$holds, because in this case every zero
of ${\cal E}_i(M)$ is also an isolated point of finite
multiplicity. By the Hellmann--Feynman theorem \cite{FHT}, the
derivative of a given ${\cal E}_i(M)$ with respect~to $M$ is
identical to the expectation value over the associated eigenstate
$|i\rangle$ (taken as normalized, $\langle i|i\rangle=1,$ for
brevity of notation) of the derivative of the Hamiltonian ${\cal
H}$ with respect to~$M$:\begin{equation}\frac{{\rm d}{\cal
E}_i}{{\rm d}M}(M)=\left\langle i\left|\frac{\partial{\cal
H}}{\partial M}\right|i\right\rangle.\label{Eq:HFT}\end{equation}

By construction, {\em all\/} the functions $h(p)$ in the
differential equation (\ref{Eq:2ODE}), and thus~{\em all\/}~our
auxiliary potentials $U(p),$ and, consequently, {\em all\/} our
Hamiltonians ${\cal H}$ exhibit a very simple, that is, a linear
dependence on the bound-state mass $M.$ The derivatives with
respect to~$M$\begin{equation}\frac{\partial{\cal H}}{\partial
M}=\frac{\partial U}{\partial M}=\frac{\partial h}{\partial M}
\label{Eq:H'}\end{equation} are summarized, for the Lorentz
structures of interaction kernels still of interest, in
Table~\ref{Tab:U'}. According to this, for precisely those choices
of the sign of the harmonic-oscillator coupling $a$ for which the
above ``infinitely-rising-potential theorem'' was found to be
applicable [viz., for $a<0$ in the case of kernels of {\em
Lorentz-scalar structure}, $\Gamma\otimes\Gamma=1\otimes1,$ and
for $a>0$ in the case of kernels of time-component Lorentz-vector
structure, $\Gamma\otimes\Gamma=\gamma^0\otimes\gamma^0,$
Lorentz-vector structure,
$\Gamma\otimes\Gamma=\gamma_\mu\otimes\gamma^\mu,$ and BJK
\cite{BJK73,Gross91} structure, $\Gamma\otimes\Gamma=\frac{1}{2}\,
(\gamma_\mu\otimes\gamma^\mu+\gamma_5\otimes\gamma_5-1\otimes1)$],
the derivatives (\ref{Eq:H'}), and thus the derivatives with
respect to $M$ of all eigenvalues ${\cal E}_i(M),$ are all
negative definite: The associated spectra of bound-state masses
$M$ will be entirely~discrete.

\begin{table}[ht]\caption{Derivative of our auxiliary
Hamiltonian operators ${\cal H}\equiv-\Delta+U$ with respect to
the bound-state mass $M$ entering in the effective potential
$U(p;M)$ as a parameter, for Lorentz structures
$\Gamma\otimes\Gamma$ entailing differential operators in the {\em
reduced instantaneous Bethe--Salpeter equation\/}
(\ref{Eq:RIBSEWEP}) with configuration-space harmonic-oscillator
interactions $V(r)=a\,r^2$
($a\ne0$).}\label{Tab:U'}\begin{center}\begin{tabular}{ccc}
\hline\hline&&\\[-1.5ex]\multicolumn{1}{c}{Lorentz structure}&
\multicolumn{1}{c}{$\Gamma\otimes\Gamma$}&
\multicolumn{1}{c}{$\displaystyle\frac{\partial{\cal H}}{\partial
M}=\frac{\partial U}{\partial M}=\frac{\partial h}{\partial
M}$}\\[2.5ex]\hline\\[-1.5ex]Lorentz scalar&$1\otimes1$&
$\displaystyle\frac{E^2(p)}{Z^2(p)\,a\,m^2(p)}$\\[2ex]
\begin{tabular}{c}time-component\\Lorentz vector\end{tabular}&
$\gamma^0\otimes\gamma^0$&$\displaystyle-\frac{1}{Z^2(p)\,a}$\\[2ex]
Lorentz vector&$\gamma_\mu\otimes\gamma^\mu$&
$\displaystyle-\frac{E^2(p)}{Z^2(p)\,a\,[E^2(p)+p^2]}$\\[2ex]BJK
\cite{BJK73,Gross91}&$\frac{1}{2}
\left(\gamma_\mu\otimes\gamma^\mu+\gamma_5\otimes\gamma_5-1\otimes1\right)$
&$\displaystyle-\frac{1}{Z^2(p)\,a}$
\\[2.5ex]\hline\hline\end{tabular}\end{center}\end{table}

Under the conditions discussed above, all effective potentials
$U(p)$ derived in Sec.~\ref{Sec:TSE} and thus all associated
Schr\"odinger Hamiltonians ${\cal H}$ are for any given value of
the parameter $M$ bounded from below. Accordingly, for each
Lorentz structure $\Gamma\otimes\Gamma$ under consideration the
spectrum of auxiliary eigenvalues ${\cal E}_i(M)$ is bounded from
below. The derivatives (\ref{Eq:HFT}) of the functions ${\cal
E}_i(M)$ proved to be negative definite. Therefore, the zero of
the ``lowest'' of all the trajectories ${\cal E}_i(M)$ defines a
lower bound on the spectrum of bound-state mass~eigenvalues.

Altogether, the sequence of findings of this section forms the
basis of the firm conviction that all bound states encountered in
any actual evaluation of the reduced \mbox{exact-propagator}
instantaneous Bethe--Salpeter equation (\ref{Eq:RIBSEWEP}) with
harmonic-oscillator interaction are stable.

\section{Summary, Conclusions, and Outlook}\label{Sec:SCO}
Motivated by instabilities observed
\cite{Parramore95,Parramore96,Olsson95,Uzzo99} for the solutions
of the Salpeter equation~with confining interaction, we
investigated the stability of the solutions of the
three-dimensional reduction of the Bethe--Salpeter equation
proposed in Ref.~\cite{Lucha05:IBSEWEP} to retain exact
propagators. Summarizing our findings, for each Lorentz structure
analyzed the solutions of our reduced {\em exact-propagator\/}
instantaneous Bethe--Salpeter equation (\ref{Eq:RIBSEWEP}) with
pure harmonic-oscillator interactions, which enter by means of
Eq.~(\ref{Eq:HOI}), exhibit very desirable characteristic
features:\begin{itemize}\item The main result is that, indeed, all
bound-state masses are discrete.\item For their spectra, we could
show: they are bounded from below.\item For all bound states, we
were able thus to prove that they are stable.\end{itemize}It goes
without saying that, in spite of some mainly technical
complications to be overcome
\cite{Lucha07:HORSE,Lucha07:StabOSS-QCD@Work07,Lucha07:SSSECI-Hadron07},
a similar stability discussion may be envisaged for the (full)
Salpeter equation \cite{Lucha:HOSE}.

\section*{Acknowledgements}W.~L.\ would like to sincerely thank
Bernhard Baumgartner for an enlightening discussion.


\small\begin{thebibliography}{30}
\bibitem{BSE}E.~E.~Salpeter and H.~A.~Bethe, Phys.~Rev.~{\bf 84}
(1951) 1232.
\bibitem{SE}E.~E.~Salpeter, Phys.~Rev.~{\bf 87} (1952) 328.
\bibitem{Lagae92}J.-F.~Laga\"e, Phys.~Rev.~D {\bf 45} (1992) 305.
\bibitem{Olsson95}M.~G.~Olsson, S.~Veseli, and K.~Williams,
Phys.~Rev.~D {\bf 52} (1995) 5141, hep-ph/9503477.
\bibitem{Olsson96}M.~G.~Olsson, S.~Veseli, and K.~Williams,
Phys.~Rev.~D {\bf 53} (1996) 504, hep-ph/9504221.
\bibitem{Lucha00:IBSEm=0}W.~Lucha, K.~Maung Maung, and
F.~F.~Sch\"oberl, Phys.~Rev.~D {\bf 63} (2001) 056002,
hep-ph/0009185.
\bibitem{Lucha00:IBSE-C4}W.~Lucha, K.~Maung Maung, and
F.~F.~Sch\"oberl, in: Proceedings of the International Conference
on {\em Quark Confinement and the Hadron Spectrum IV}, edited by
W.~Lucha and~K.\ Maung Maung (World Scientific, New
Jersey/London/Singapore/Hong Kong, 2002), p.~340, hep-ph/0010078.
\bibitem{Lucha00:IBSEnzm}W.~Lucha, K.~Maung Maung, and
F.~F.~Sch\"oberl, Phys.~Rev.~D {\bf 64} (2001) 036007,
hep-ph/0011235.
\bibitem{Lucha01:IBSEIAS}W.~Lucha and F.~F.~Sch\"oberl,
Int.~J.~Mod.~Phys.~A {\bf 17} (2002) 2233, hep-ph/0109165.
\bibitem{Lucha05:IBSEWEP}W.~Lucha and F.~F.~Sch\"oberl, J.~Phys.~G:
Nucl.~Part.~Phys.\ {\bf 31} (2005) 1133, hep-th/0507281.
\bibitem{Lucha05:EQPIBSE}Li Z.-F., W.~Lucha, and F.~F.~Sch\"oberl,
Mod.~Phys.~Lett.~A {\bf 21} (2006) 1657, hep-ph/0510372.
\bibitem{Parramore95}J.~Parramore and J.~Piekarewicz, Nucl.~Phys.~A
{\bf 585} (1995) 705, nucl-th/9402019.
\bibitem{Parramore96}J.~Parramore, H.-C.~Jean, and J.~Piekarewicz,
Phys.~Rev.~C {\bf 53} (1996) 2449, nucl-th/9510024.
\bibitem{Uzzo99}M.~Uzzo and F.~Gross, Phys.~Rev.~C {\bf 59} (1999)
1009, nucl-th/9808041.
\bibitem{BJK73}M.~B\"ohm, H.~Joos, and M.~Krammer, Nucl.~Phys.~B
{\bf 51} (1973) 397.
\bibitem{Gross91}F.~Gross and J.~Milana, Phys.~Rev.~D {\bf 43}
(1991) 2401.
\bibitem{Henriques76}A.~B.~Henriques, B.~H.~Kellett, and
R.~G.~Moorhouse, Phys.~Lett.~B {\bf 64} (1976) 85.
\bibitem{Jacobs87}S.~Jacobs, M.~G.~Olsson, and C.~J.~Suchyta III,
Phys.~Rev.~D {\bf 35} (1987) 2448.
\bibitem{Gara89}A.~Gara, B.~Durand, L.~Durand, and L.~J.~Nickisch,
Phys.~Rev.~D {\bf 40} (1989) 843.
\bibitem{Gara90}A.~Gara, B.~Durand, and L.~Durand, Phys.~Rev.~D
{\bf 42} (1990) 1651; {\em ibid.} {\bf 43} (1991) 2447 (erratum).
\bibitem{Lucha92C}W.~Lucha, H.~Rupprecht, and F.~F.~Sch\"oberl,
Phys.~Rev.~D {\bf 45} (1992) 385.
\bibitem{Lucha07:HORSE}Z.-F. Li, W.~Lucha, and F.~F.~Sch\"oberl,
Phys.~Rev.~D (in press), arXiv:0707.3202 [hep-ph].
\bibitem{Lucha07:StabOSS-QCD@Work07}W.~Lucha and F.~F.~Sch\"oberl,
in: Proceedings of {\em QCD@Work 2007: International Workshop on
Quantum Chromodynamics: Theory and Experiment}, edited by
P.~Colangelo, D.~Creanza, F.~De Fazio, R.~A.~Fini, E.~Nappi, and
G.~Nardulli, AIP Conf.~Proc.\ (AIP, New York, 2007), Vol.~964,
p.~318, arXiv:0707.1440 [hep-ph].
\bibitem{Lucha07:SSSECI-Hadron07}W.~Lucha and F.~F.~Sch\"oberl,
preprint HEPHY-PUB 856/07 (2007), arXiv:0711.1736~[hep-ph], to
appear in the Proceedings of {\em Hadron 07, XIIth International
Conference on Hadron Spectroscopy}, Frascati (Rome), Italy,
October 8--13, 2007.
\bibitem{Maris97}P.~Maris and C.~D.~Roberts, Phys.~Rev.~C {\bf 56}
(1997) 3369, nucl-th/9708029.
\bibitem{Abramowitz}{\em Handbook of Mathematical Functions},
edited by M.~Abramowitz and I.~A.~Stegun (Dover, New York, 1964).
\bibitem{RS4}M.~Reed and B.~Simon, {\em Methods of Modern
Mathematical Physics IV: Analysis of Operators\/} (Academic Press,
New York, 1978).
\bibitem{FHT}H.~Hellmann, Acta Physicochim.\ URSS {\bf 1}
(1934/1935) 913; {\em ibid.} {\bf 4} (1936) 225; {\em
Einf\"uhrung~in die Quantenchemie\/} (F.~Deuticke, Leipzig/Wien,
1937), p.~285;

R.~P.~Feynman, Phys.~Rev.~{\bf 56} (1939) 340.
\bibitem{Lucha:HOSE}W.~Lucha and F.~F.~Sch\"oberl (in
preparation).
\end{thebibliography}
\end{document}